\documentclass[graybox]{svmult}

\usepackage{mathptmx}       
\usepackage{helvet}         
\usepackage{courier}        
\usepackage{type1cm}        
\usepackage{makeidx}         
\usepackage{graphicx}        
\usepackage{multicol}        
\usepackage[bottom]{footmisc}

\makeindex             

\begin{document}

\title*{Axial Quasi-normal modes of Neutron Stars with Exotic Matter} 
\author{J. L. Bl\'azquez-Salcedo, L. M. Gonz\'alez-Romero and
  F. Navarro-L\'erida} 
\institute{J. L. Bl\'azquez-Salcedo and L. M. Gonz\'alez-Romero \at Depto. F\'{\i}sica Te\'orica II,
  Facultad de  Ciencias F\' \i sicas, Universidad Complutense de Madrid,
  28040-Madrid, Spain \email{joseluis.blazquez@fis.ucm.es, mgromero@fis.ucm.es}
\and F. Navarro-L\'erida \at Depto. F\'{\i}sica \'Atomica, Molecular y Nuclear, Facultad de  Ciencias F\'
\i sicas, Universidad Complutense de Madrid, 28040-Madrid, Spain \email{fnavarro@fis.ucm.es}}

%
%
\maketitle

\abstract{
We investigate the axial w quasi-normal modes of neutron stars for
18 realistic equations of  
state in order to study the influence of hyperons and quarks on the modes. 
The study has been developed with a new method based on Exterior
Complex Scaling with variable angle,
which allow us to generate pure
outgoing 
quasi-normal modes. A complete study of
the junction conditions has been done. 
We have obtained that 
w-modes can be used to 
distinguish between neutron stars with
exotic matter and without exotic matter for compact enough stars.
}

\section{Quasi-normal modes Formalism and Numerical Method}
\label{sec:1}

We will consider a spherical and static star. The matter inside of it
is considered to be a perfect fluid. Following the original 
papers (see reviews \cite{Kokkotas_Schmidt1999,Nollert1999}), we
make perturbations over 
the spherical static metric and the stress-energy tensor, taking into 
account only the axial perturbations. After some algebra, it
can be seen that the perturbations satisfy the well-known Regge-Wheeler equation
\cite{PhysRevLett.24.737}, both inside and outside the star. 
The eigen-frequency of the axial mode is a complex number $\omega =
\omega_{\Re} + i\omega_{\Im}$.
Inside the star an equation of state must be provided, so in general also the
static configuration must be solved numerically. Outside the star the metric
is known (Schwarszchild) and only the perturbation must be integrated.

We are only interested in purely outgoing waves. In
general a solution of the Regge-Wheeler equation will be 
a composition of incoming and outgoing oscillating waves. Because the outgoing
wave diverges towards infinity, the purely outgoing quasi-normal mode 
condition could only be imposed as a behavior far enough from the star, but
every small numerical error in the imposition of this 
behavior will be 
amplified as we approach the border of the star, resulting in a mixture of
outgoing with ingoing waves. Note also that in general the exterior solution
will oscillate infinitely towards infinity. We have developed the following
method, based on the Colsys package \cite{colsys1979}, to deal with these difficulties. We make use of previously well known techniques and new ones. 

\textbf{\emph{Exterior solution:}} We study the phase function (logarithmic derivative of the Regge-Wheeler function)
, which does not oscillate. Hence, the
differential equation outside the star is reduced to a Riccati equation and we
can compactify the radial variable. The boundary condition must grant the
outgoing wave behavior.
In order to impose a constringent enough condition, we make use of Exterior
Complex Scaling method \cite{BlazquezSalcedo:2012pd} with variable angle, where the integration coordinate is considered to be a complex variable. The principal advantage with respect other methods is that in principle no assumption on the imaginary part (i.e. damping time) of the quasi-normal mode is done.

\textbf{\emph{Interior solution:}} 
The interior part of the solution is integrated numerically.
As we want to obtain realistic configurations, we implement the equations of
state in two different ways: 1) A piece-wise polytrope
approximation, done by Read et al \cite{Jocelyn2009}, in which the equation
of state is approximated by a polytrope in different density-pressure
intervals. 2) A piece-wise monotone 
cubic Hermite interpolation satisfying local thermodynamic conditions
. 

We generate two independent
solutions inside of the star for the same static configuration. These two
solutions must be combined to match the exterior solution with the appropriate
junction conditions. We use Darmois conditions (continuity of the fundamental
forms of the matching hypersurface). This formulation allow us to introduce
surface 
layers of energy density on the border of the star
, that  
allow us to approximate the exterior \textit{crust} as a thin layer enveloping
the core. 

\textbf{\emph{Determinant method:}} The junction conditions can be used to
construct what we call the 
\emph{determinant method}: 
We construct a matrix in terms of the
derivatives of the Regge-Wheeler function whose determinant must be zero only
if the matching conditions are fulfilled, i.e., when $\omega$ corresponds to a
quasi-normal mode for the static configuration integrated. The matrix is calculated using both independent solutions in the interior of the star together with the exterior phase function.

This method has been successfully extended to study polar modes of realistic neutron stars. These results will be presented elsewhere.

\section{Numerical Results}
We have made several tests on our method successfully reproducing data
from previous works for axial modes. As an example, we reproduce the results from \cite{Samuelsson2007} with a precision of $10^{-7}$. In this section we 
will present our results for new realistic EOS. 
Using the parametrization presented by Read et all \cite{Jocelyn2009}, we can
study the  
34 equations of
state they considered. We have used, following their notation,
SLy, APR4, BGN1H1, GNH3, H1, H4, ALF2, ALF4. After the recent measurement of the $1.97 M_\odot$ for the pulsar PSR
J164-2230 \cite{Demorest2010}, several exotic matter EOS have been proposed
satisfying this condition. We have
considered the following ones using the cubic Hermite interpolation:   two EOS
presented by
Weissenborn et al with hyperons in \cite{Weissen1},  we call them  WCS1 y WCS2;
three EOS presented by  Weissenborn et al with quark matter in
\cite{Weissen2}, we call them WSPHS1, WSPHS2,  WSPHS3; four EOS
presented by L. Bonanno and A. Sedrakian in \cite{Sedrakian}; we call them BS1,
BS2, BS3, BS4; and one EOS presented by Bednarek et al in \cite{Bednarek}, we
call it BHZBM.

 \begin{figure}
\begin{center}
 \includegraphics[angle=-90,width=0.46\textwidth]{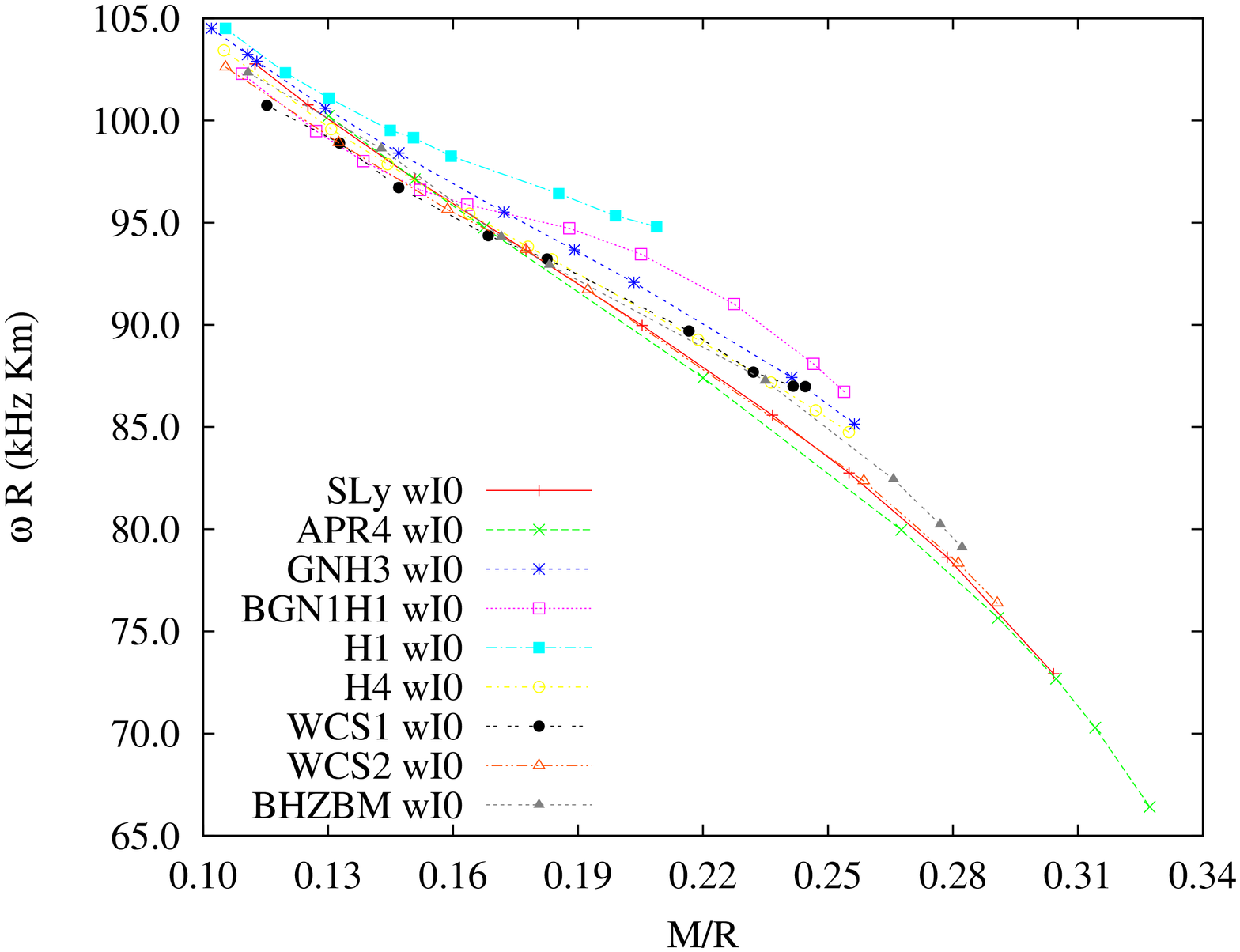} 
 \includegraphics[angle=-90,width=0.46\textwidth]{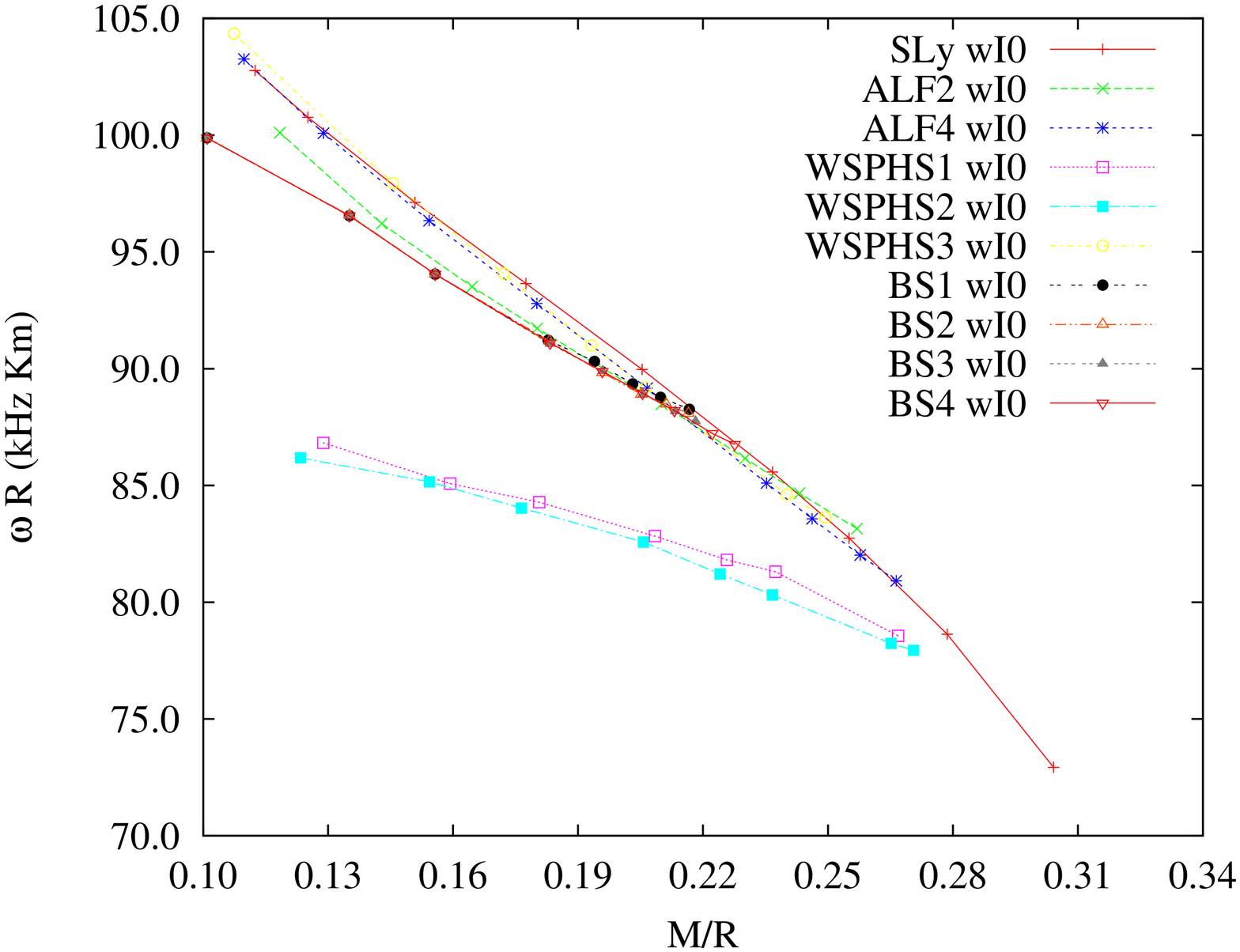} 
 \includegraphics[angle=-90,width=0.46\textwidth]{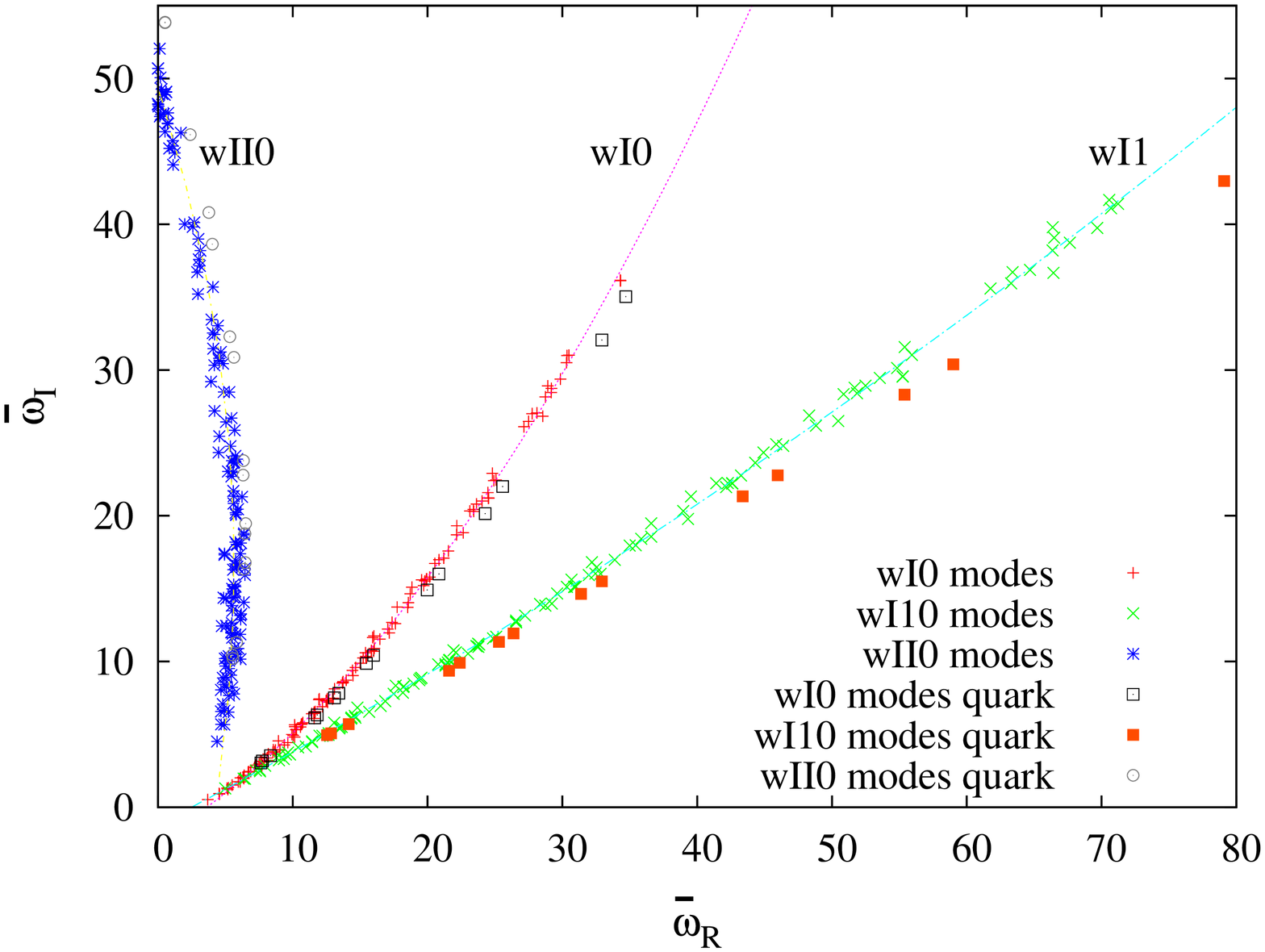} 
 \includegraphics[angle=-90,width=0.46\textwidth]{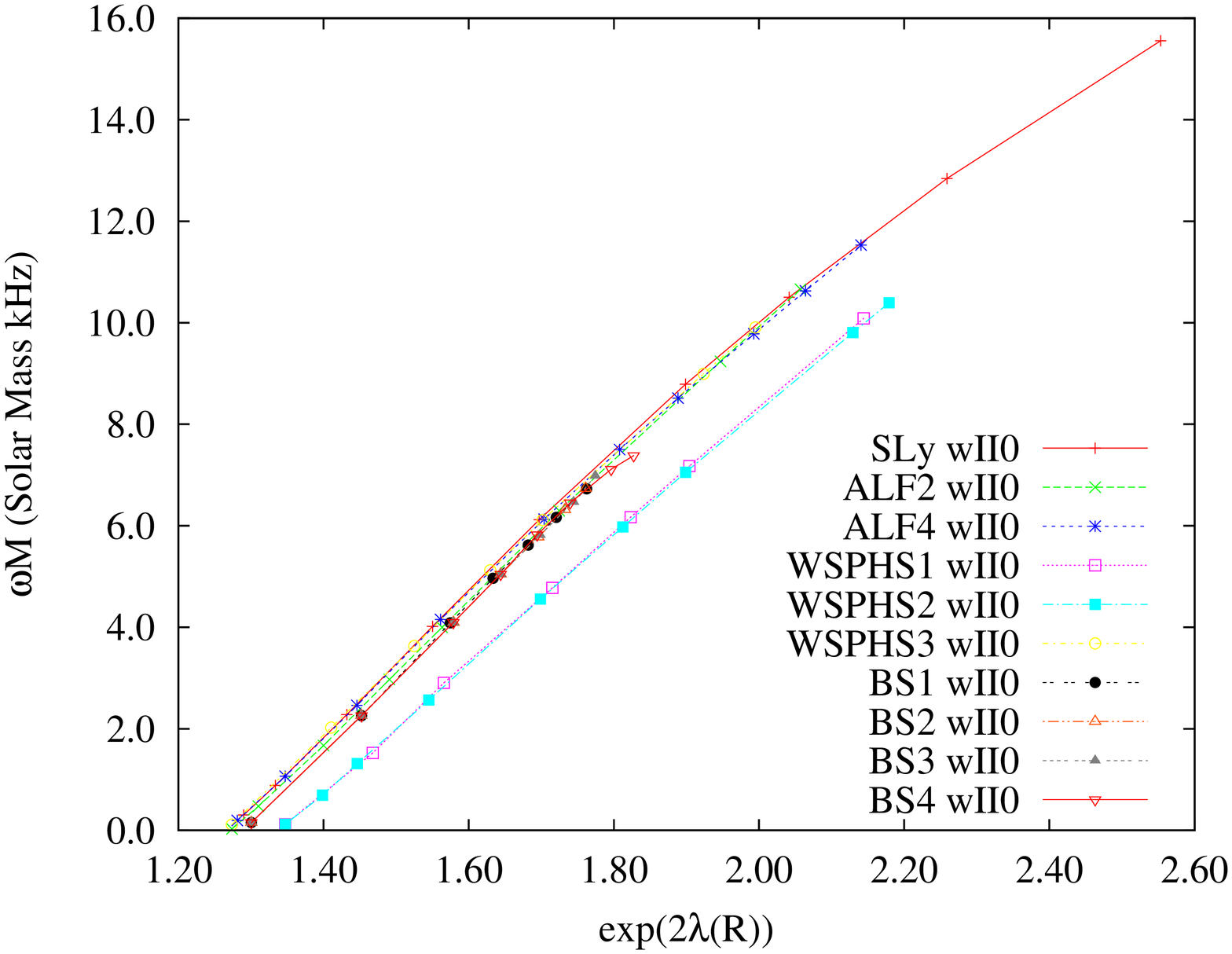}
\end{center}
 \caption{Top: Scaled frequency of the fundamental wI mode vs M/R for
   stars 
   with hyperon (left) and quark matter (right). Bottom left: All
   w-modes in units of the central pressure for all 
    EOS and different central pressures. Bottom right: Scaled frequency of the fundamental wII mode vs $e^{2\lambda(R)}$.}
 \label{eos_wI0_real_pc}
 \end{figure}

Empirical relations between the frequency and damping time of 
quasi-normal modes and the compactness of the star can be useful in order to use future observations of gravitational waves to estimate  the
mass and the radius of the neutron star, as well as  
 to discriminate between different families of equations of state. In top of Figure \ref{eos_wI0_real_pc}  we present  the
frequency of the fundamental mode scaled to the radius of each
configuration. The
softest equations of state that include hyperon matter, H1 and BGN1H1, present a quite
different behavior than the rest of EOS considered. Nevertheless,
as the detection of the recent $2 M_{\odot}$ pulsar suggest, these two
particular EOS cannot be realized in nature.

Another exception is found in pure quark matter stars (WSPHS1-2 EOS). Their behavior is clearly differentiated from the rest because of the 
different layer structure found at the exterior of the star.

In general, for hyperon matter EOS and hybrid stars, we obtain linear relations between the scaled frequency and the compactness
.
These relations could be used, applying the technique from
\cite{Kokkotas_asteroseismology_1998}, to measure the radius of the neutron
star and constrain the equation of state.   

We plot at the bottom left of Fig. \ref{eos_wI0_real_pc} a new
phenomenological relation between the real part and the imaginary part of the  
 frequency of the w quasi-normal modes valid for all the EOS. We plot
$\bar \omega_{R} =2 \pi  \frac{1}{\sqrt{p_c(cm^{-2})}}\frac{10^3}{c} \omega(Khz)$ and $\bar \omega_{I}= \frac{1}{\sqrt{p_c(cm^{-2})}}\frac{10^6}{c} \frac{1}{\tau(\mu  s)}$.
Although the empirical relation between $\bar \omega_{R}$ and $\bar \omega_{I}$ is quite independent of the EOS, the parametrization of the curve is EOS dependent. So a possible application of this empirical relation is the following. If the frequency $\omega(Khz)$ and 
the damping time $\tau(\mu s)$ are known, we can parametrize a line defining $\bar \omega_{R}$ and $\bar \omega_{I}$ with parameter $p_c$ using the observed frequency and damping time. The
crossing point of this line with the empirical relation presented in the bottom left of Fig. \ref{eos_wI0_real_pc} gives us an estimation of the central pressure
$p_c$ independent of the EOS. Now, we can check which EOS is compatible with this $p_c$, i.e., which one  
have the measured wI0 mode near the crossing point for the estimated central
pressure. Hence, this method could be used to constrain the equation of state. Note that if 
mass and radius are already measured, we would have another filter to impose to the EOS. 

 Also, the precision of our algorithm allows us to construct  explicitly the
 universal low compactness limiting configuration for fundamental wII modes
 (bottom right of Fig. \ref{eos_wI0_real_pc}) around $M/R=0.106$ for
 which the fundamental wII mode vanishes \cite{Wen2009}.

We also study the impact of the
core-\textit{crust} transition pressure on the quasi-normal mode spectrum. We obtain
that variations of the transition pressure from  $10^{32}
dyn/cm^{2}$ to $10^{33}
dyn/cm^{2}$ affect the frequency and damping time order $0.1\%$.

\bibliographystyle{spphys}
\bibliography{axial_proc_bib}

\end{document}